\documentclass[twocolumn,showpacs,preprintnumbers,amsmath,amssymb,floatfix]{revtex4}
\usepackage{graphicx}
\usepackage{dcolumn}
\usepackage{bm}

\begin{document}

\draft
\title{
\vspace{-15mm}
\begin{flushright}
\footnotesize{Resubmitted to Physical Review Letters \\
January 2002}
\end{flushright}
Statically Transformed Autoregressive Process and Surrogate
  Data Test for Nonlinearity}

\author{D. Kugiumtzis}

\email{dkugiu@gen.auth.gr}

\homepage{http://users.auth.gr/dkugiu}

\affiliation{Department of Mathematical and Physical Sciences, 
Polytechnic School, \\ 
Aristotle University of Thessaloniki, Thessaloniki 54006, Greece}

\begin{abstract}
The key feature for the successful implementation of the surrogate data
test for nonlinearity on a scalar time series is the generation of 
surrogate data that represent exactly the null hypothesis (statically
transformed normal stochastic process), i.e. they possess the sample 
autocorrelation and amplitude distribution of the given data.
A new conceptual approach and algorithm for the generation of
surrogate data is proposed, called {\em statically transformed
  autoregressive process} (STAP).
It identifies a normal autoregressive process and a monotonic static 
transform, so that the transformed realisations of this process fulfill 
exactly both conditions and do not suffer from bias in autocorrelation 
as the surrogate data generated by other algorithms. 
The appropriateness of STAP is demonstrated with simulated and  
real world data.
\end{abstract}


\maketitle

The surrogate data test for nonlinearity has been widely used in real 
applications in order to establish statistically the existence of
nonlinear dynamics and justify the use of nonlinear tools in the 
analysis of time series \cite{Kantz97,Diks00}.
The most general null hypothesis $\mbox{H}_0$ of this test so far is
that the examined time series ${\mathbf x} = [x_1, \ldots,x_n]'$ is
a realisation of a normal (linear) process ${\mathbf s} \!=\! [s_1,
\ldots,s_n]'$ undergoing a possibly nonlinear static transform $h$,
i.e. $x_i \!=\! h(s_i)$, $i\!=\!1,\ldots,n$.
To test the null hypothesis, an estimate $q$ from a nonlinear method
applied to the original data, $q_0$, is compared to the estimates,
$q_1,\ldots,q_M$ on $M$ surrogate time series representing
the null hypothesis \cite{Schreiber99b,Kugiumtzis01a}. 
A  properly designed surrogate time series ${\mathbf z}$ should
possess the same autocorrelation as the original data, 
$r_z(\tau) = r_x(\tau)$ for a range of lags $\tau$, and the same 
amplitude distribution, $F_z(z_i) = F_x(x_i)$ ($F_x(x_i)$ is the 
cumulative density function (cdf) of $x_i$), and be otherwise random. 

It has been reported that erroneous results are likely to occur mainly 
due to the insufficiency of the algorithms to generate surrogate data 
that preserve the original linear correlations \cite{Kugiumtzis00}. 
The prominent algorithm of amplitude adjusted Fourier transform (AAFT)
\cite{Theiler92}, used in most applications so far, is built based on 
the assumption of monotonicity of $h$. 
When $h$ is not monotonic, the AAFT algorithm is found to 
favour the rejection of $\mbox{H}_0$ due to the mismatch of the
original linear correlations \cite{Kugiumtzis99}. 
The iterated AAFT (IAAFT) algorithm improves the match of the 
autocorrelation of AAFT \cite{Schreiber96}, but with about the same 
accuracy for all the surrogates, so that the small variance, combined 
with the small bias, may be another cause for false rejections in some 
cases \cite{Kugiumtzis99}. 
Another algorithm making use of simulation annealing
seems to perform similarly to IAAFT \cite{Schreiber98}.  
Recently, a correction of AAFT (CAAFT) that results in 
unbiased match of the linear correlations was proposed in 
\cite{Kugiumtzis00d}. 

In this paper, we develop further the somehow profound rationale 
of the CAAFT algorithm and formulate a new conceptual approach for the 
generation of surrogate data consistent with $\mbox{H}_0$, which we
then solve analytically.  
The new algorithm, called STAP, generates the surrogate data as 
realisations of a suitable {\em statically transformed autoregressive 
process} (STAP), i.e. the process under $\mbox{H}_0$ is designed 
as a static transform of a suitable normal process. 

The main idea behind the STAP algorithm is that for {\em any} stationary 
process ${\mathbf X} = [X_1,X_2, \ldots]'$, for which we have finite scalar 
measurements ${\mathbf x}$, there is a scalar {\em linear} stochastic 
process ${\mathbf Z}$ with the same autocorrelation $\rho_X$ and marginal 
cdf $\Phi_X$ as for the observed process ($\rho_Z(\tau)=\rho_X(\tau)$ and
$\Phi_Z(z_i)=\Phi_X(x_i)$), i.e. ${\mathbf Z}$ is a scalar ``linear
copy'' of the observed ${\mathbf X}$. 
The objective is to derive ${\mathbf Z}$ through a static monotonic
transform $g$ on a scalar normal process ${\mathbf U}$ with a proper
autocorrelation $\rho_U$, i.e. $Z_i=g(U_i)$. 
Thus $g$ and $\rho_U$ have to be properly selected, so that 
${\mathbf Z}$ has the desired properties.
In practice, the surrogate data set is a finite realisation
${\mathbf z} = [z_1, \ldots,z_n]'$ of the process ${\mathbf Z}$ and
$g$ and $\rho_U$ are estimated based solely on ${\mathbf x}$.
Note that $g$ and ${\mathbf U}$ are in general different from $h$ and 
${\mathbf S}$, respectively, of $\mbox{H}_0$ (they are the same if $h$ 
is monotonic \cite{Kugiumtzis00d}).
Thus with this approach $\mbox{H}_0$ can be formulated more generally
that the time series is generated by a linear stochastic process.

Let $\Phi_0$ be the marginal cdf of a standard normal process 
${\mathbf U}$.
A suitable choice for $g$, so that $\Phi_Z(z_i)=\Phi_X(x_i)$, is 
defined as \cite{Papoulis91}  
\begin{equation}
  Z_i=g(U_i) = \Phi_X^{-1}(\Phi_0(U_i)),
  \label{eq:g}
\end{equation}
where $g$ is monotonic by construction. 
Assuming that $\Phi_X(x_i)$ is continuous and strictly increasing 
and that $-1 < \rho_U < 1$, which are both true for 
all practical purposes, there is a function $\phi$ depending on $g$, 
such that $\rho_Z=\phi(\rho_U)$ for any lag $\tau$ 
\cite{Johnson72,DerKiureghian86}.  
If $g$ has an analytic form, then it may be possible to find an analytic 
expression for $\phi$ as well. 
In that case, given that $\rho_X$ is known and by setting 
$\rho_Z\equiv\rho_X$, one can invert $\phi$
to find $\rho_U = \phi^{-1}(\rho_X)$, if $\phi^{-1}$ exists.

In general, the function $g$, as defined in eq.(\ref{eq:g}), does 
not have an analytic form because $\Phi_X$ is not known analytically, 
but it can be approximated by an analytic function, e.g. a polynomial
$p_m$ of degree $m$,
\begin{equation} 
  Z_i = g(U_i) \simeq p_m(U_i) = a_0 + \sum_{j=1}^m a_j U_i^j.
  \label{eq:pm}
\end{equation}
Low degree polynomials have been used to approximate such transforms 
\cite{Elphinstone83,Hawkins94}. 
Then using the definition for the autocorrelation, the approximate
expression for $\phi$ reads  
\begin{equation}
  \rho_X = \phi(\rho_U) = \frac{\sum_{s=1}^m \sum_{t=1}^m a_s a_t
    (\mu_{s,t} - \mu_s \mu_t)}{\sum_{s=1}^m \sum_{t=1}^m a_s a_t
    (\mu_{s+t} - \mu_s \mu_t)},
  \label{eq:phi}
\end{equation}
where an arbitrary lag $\tau$ is implied as argument for the
autocorrelations, $\mu_s$ is the $s$th central moment of $U_i$ being
$\mu_{2k+1}=0$, $\mu_{2k} =1\cdot3\cdots(2k-1)$, $k\geq 0$, and
$\mu_{s,t}$ is the $s$th-$t$th central joint moment of the bivariate
standard normal distribution of $(U_i,U_{i-\tau})$, defined as
follows \cite{Hutchinson90}
\begin{eqnarray*} 
  \mu_{2k,2l} & = & \frac{(2k)!(2l)!}{2^{k+l}} \sum_{j=0}^{\nu}
  \frac{(2\rho_U)^{2j}}{(k-j)!(l-j)!(2j)!} \\
  \mu_{2k+1,2l+1} & = & \frac{(2k+1)!(2l+1)!}{2^{k+l+1}} \sum_{j=0}^{\nu}
  \frac{(2\rho_U)^{2j+1}}{(k-j)!(l-j)!(2j+1)!} \\
  \mu_{i,j} & = & 0 \quad \mbox{if} \quad k+l = \mbox{odd},
\end{eqnarray*}
where $\nu=\min(k,l)$.
By substituting the expression for the moments in eq.(\ref{eq:phi}),
the expression for $\phi$ can be brought to a polynomial form of the
same order $m$ 
\begin{equation}
  \rho_X = \phi(\rho_U) = \sum_{j=1}^m c_j \rho_U^j,
  \label{eq:phi2}
\end{equation}
where the vector of coefficients ${\mathbf c}=[c_1,\ldots,c_m]'$ is
expressed only in terms of ${\mathbf a}=[a_1,\ldots,a_m]'$ (the 
expressions are rather involved and therefore not presented here).
Simpler expressions can be derived using the Tchebycheff-Hermite 
polynomials \cite{Bhatt64}. 
Thus an analytic expression for $\rho_U$ is possible if
eq.(\ref{eq:phi2}) can be solved with respect to $\rho_U$. 
From our simulations, we conjecture that if $g$ is monotonic then
$\phi$ is also monotonic in $[-1,1]$. 
Then $\phi^{-1}$ exists and a unique solution for $\rho_U$ can be
found from eq.(\ref{eq:phi2}). 
The proper standard normal process ${\mathbf U}$ is completely defined
by $\rho_U$ and applying the
transform $g$ of eq.(\ref{eq:g}) to the components of ${\mathbf U}$,
the ``linear copy'' ${\mathbf Z}$ of the given process ${\mathbf X}$ is
constructed. 

Note that the solution for $\rho_U$ is given analytically from the 
polynomial approximation of $g$ and it requires only the
knowledge of the coefficients ${\mathbf a}$ of the polynomial and the
autocorrelation $\rho_X$. 

In practice, we operate with a single time series ${\mathbf x}$ rather
than a process ${\mathbf X}$ and with the sample estimates $F_x$
and $r_x$ for $\Phi_X$ and $\rho_X$, respectively. 
The steps of the algorithm are as follows:
\begin{enumerate}
  \item Estimate the vector of coefficients
    ${\mathbf a}=[a_1,\ldots,a_m]'$ of the polynomial $p_m$ from the
    graph of $x_i = F_{x}^{-1}(F_0(w_i))$, i.e. the 
    graph of ${\mathbf x}$ vs ${\mathbf w}$ after their ranks are
    matched, where ${\mathbf w}=[w_1,\ldots,w_n]'$ is standard white
    normal noise.
  \item Compute ${\mathbf c}=[c_1,\ldots,c_m]'$ for the given
    ${\mathbf a}$ from eqs.(\ref{eq:phi}--\ref{eq:phi2}). 
  \item Find $r_u$ from eq.(\ref{eq:phi2}) for the given ${\mathbf c}$
    and $r_x$ using the sample estimates $r$ instead of $\rho$. 
    The common practice is that the solution exists and it is unique.
    If this is not the case, repeat the steps 1--3 for a new ${\mathbf 
      w}$ until a unique solution is obtained.
  \item Generate a realisation ${\mathbf u}$ of a standard normal
    process with autocorrelation $r_u$. 
    We choose to do this simply by means of an autoregressive model of
    some order $p$, AR($p$). 
    The parameters ${\mathbf b}=[b_0,b_1,\ldots,b_p]'$ of AR($p$) are
    found from $r_u$ using the normal equations solved effectively by
    the Levinson algorithm \cite{Brockwell91}. 
    The AR($p$) model is run to generate ${\mathbf u}$
    \[ 
      u_{i+1} = b_0 + \sum_{j=1}^p b_j u_{i-(j-1)} + e_i, \quad e_i
        \sim \mbox{N}(0,1). 
    \] 
  \item Transform ${\mathbf u}$ to ${\mathbf z}$ by reordering
    ${\mathbf x}$ to match the rank order of ${\mathbf u}$, i.e. $z_i =
    F_x^{-1}(F_0(u_i))$.  
\end{enumerate}

Note that ${\mathbf u}$ possesses the sample normal marginal cdf $F_0$
and the proper $r_u$, so that ${\mathbf z}$ possesses $F_z =
F_x$, $r_z =r_x$, and is otherwise random, as desired.
In practice however, the equality $r_z =r_x$ is not exact and $r_z$ 
may vary substantially around $r_x$. 
Two possible reasons for this are the insufficient approximation of $g$
in step 1 and the inevitable variation of the sample autocorrelation
of the generated ${\mathbf u}$ in step 4, which decreases with the
increase of data size. 
The former is due to the limited power of polynomials in approximating 
monotonic functions and this shortcoming causes also occasional 
repetitions of the first steps of the algorithm as stated in step 3
\footnote{Other model classes may give better approximations of $g$ 
and should be investigated, but polynomials were used here to derive 
a simple analytic form for the transform of the autocorrelations.}.
The latter constitutes an inherent property of the so-called ``typical
realisation'' approach (i.e. a model is used to generate the surrogate
data) and cannot be controlled.  
However, less variation in the autocorrelation is achieved when the
AR($p$) model is optimised making the following steps, in the same
way as for the CAAFT algorithm \cite{Kugiumtzis00d}:
\begin{enumerate}
  \item Apply the algorithm presented above $K$ times and get
    ${\mathbf z}^1,\ldots,{\mathbf z}^K$ surrogate time series. 
  \item Compute $r_{z^1},\ldots,r_{z^K}$ and find the one, $r_{z^l}$
    closest to $r_x$ \footnote{We found that a robust way to achieve this is to compute the error
  $\sum_{i=1}^{\tau}(r_x(i)-r_{z^j}(i))^2$ for $j\!=\!1,\ldots,K$ and
  $\tau\!=\!1,\ldots,\tau_{\mbox{\scriptsize{max}}}$ and then select the trial
  $l$ that gives the minimum error most times, where
  $\tau_{\mbox{\scriptsize{max}}}$ minimums are totally computed, each time
  over the $K$ trials.}.
  \item Use the parameters ${\mathbf b}$ of the $l$-repetition to
    generate the $M$ surrogate data (steps 4--5 of the algorithm above). 
\end{enumerate}
The $K$ repetitions above as well as the occasional repetitions of steps
1--3 of the first part of the algorithm may slow down the algorithm 
if the time series is long, but they have no impact on the principal
function of the algorithm.  
Simply, some realisations of white noise $\mathbf w$ are discarded
in the search of the parameters ${\mathbf b}$ of the most suitable AR 
model that generates the surrogate data (through the $g$ 
transform).
 
The free parameters of the STAP algorithm are the degree $m$ of the
polynomial approximation of $g$, the order $p$ of the AR model, the
number $K$ of repetitions for the optimisation of AR($p$) and the
maximum lag $\tau_{\mbox{\scriptsize{max}}}$, used to compare
$r_{z^1},\ldots,r_{z^K}$ to $r_x$.
Usually, a small $m$ ($m \leq 10$) is sufficient. 
For $p$, there is no optimal range of values but it may vary with
the shape of $r_x$, e.g. a slowly decaying $r_x$ may be better
modelled by a larger $p$.
In all our simulations, we set $K=M=40$ and 
$\tau_{\mbox{\scriptsize{max}}} = p$. 

\begin{figure}[htb] 
\centerline{\includegraphics[width=7cm,keepaspectratio]{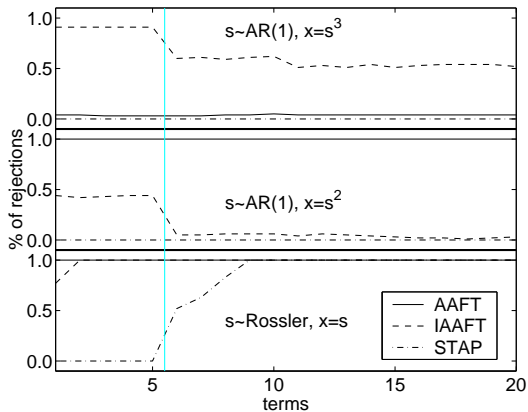}}
\caption{Percentages of rejections of $\mbox{H}_0$ using as 
discriminating statistics the fit of Volterra polynomials from $100$ 
realisations for each of the three cases in the three panels, as 
indicated. 
Three types of surrogates are used in each test as shown in the legend
(for STAP $m=5$, $p=5$).
The vertical gray line distinguishes the linear from the nonlinear 
statistics.
}  
\label{Synthetic}
\end{figure}

The proper performance and superiority of STAP over AAFT and IAAFT was 
confirmed from simulations on different toy models.
CAAFT was found to perform very similarly to STAP.
We show in Fig.~\ref{Synthetic} comparative results for AAFT, IAAFT and 
STAP for three representative synthetic systems: the cube of an AR(1) 
process ($s_{i+1} = 0.3 + 0.8 s_i + e_i, e_i \sim \mbox{N}(0,1),
x_i = s_i^3$), the square of the same AR(1) ($x_i = s_i^2$), 
both being consistent with $\mbox{H}_0$, and the $x$ variable of 
the R\"{o}ssler system \cite{Roessler76}, not consistent with 
$\mbox{H}_0$.
For each system, we generate 100 time series of $2048$ samples each
and for each realisation we generate $M=40$ surrogate data of each type. 

As discriminating statistics $q^i$ we choose the correlation
coefficient (CC) of the fit with the series of Volterra polynomials
of degree 2 and order $v = 5$.
The polynomials for the first $i$ terms, where 
$i=1,\ldots,5$, are linear and for terms $i=6,\ldots,20$ are nonlinear 
(see also \cite{Kugiumtzis00d}). 
To quantify the discrimination we use the significance $S^i=\frac{|q_0^i -
  \langle q^i \rangle|}{\sigma_q^i}$ for each polynomial of $i$ terms, 
where $q_0^i$ is the statistic $q^i$ on the original data, 
$\langle q^i \rangle$ and $\sigma_q^i$ are the mean and standard 
deviation of the statistic $q^i$ on the $M$ surrogate data. 
The null hypothesis $\mbox{H}_0$ is formally rejected at the 0.05
significance level when $S>1.96$, under the assumption of normality 
for the statistic $q$, which turns out to hold in general.   
The percentages of rejections for each of the three systems are shown 
in Fig.~\ref{Synthetic}.
Very similar results were found when deciding the rejection from 
the rank ordering of $q_0^i,q_1^i,\ldots,q_M^i$. 

\begin{figure}[htb] 
\centerline{\includegraphics[width=7cm,keepaspectratio]{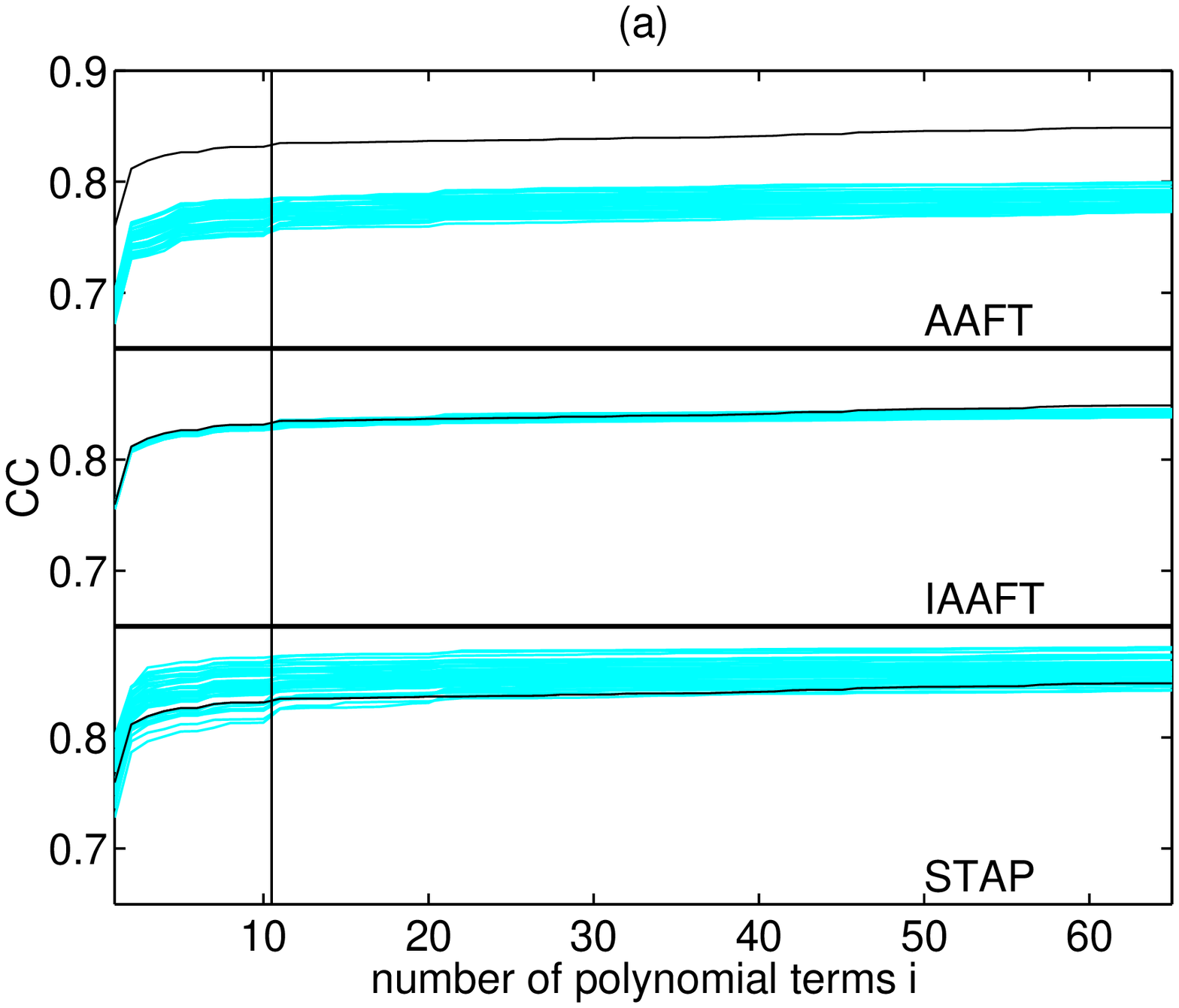}}
\centerline{\includegraphics[width=7cm,keepaspectratio]{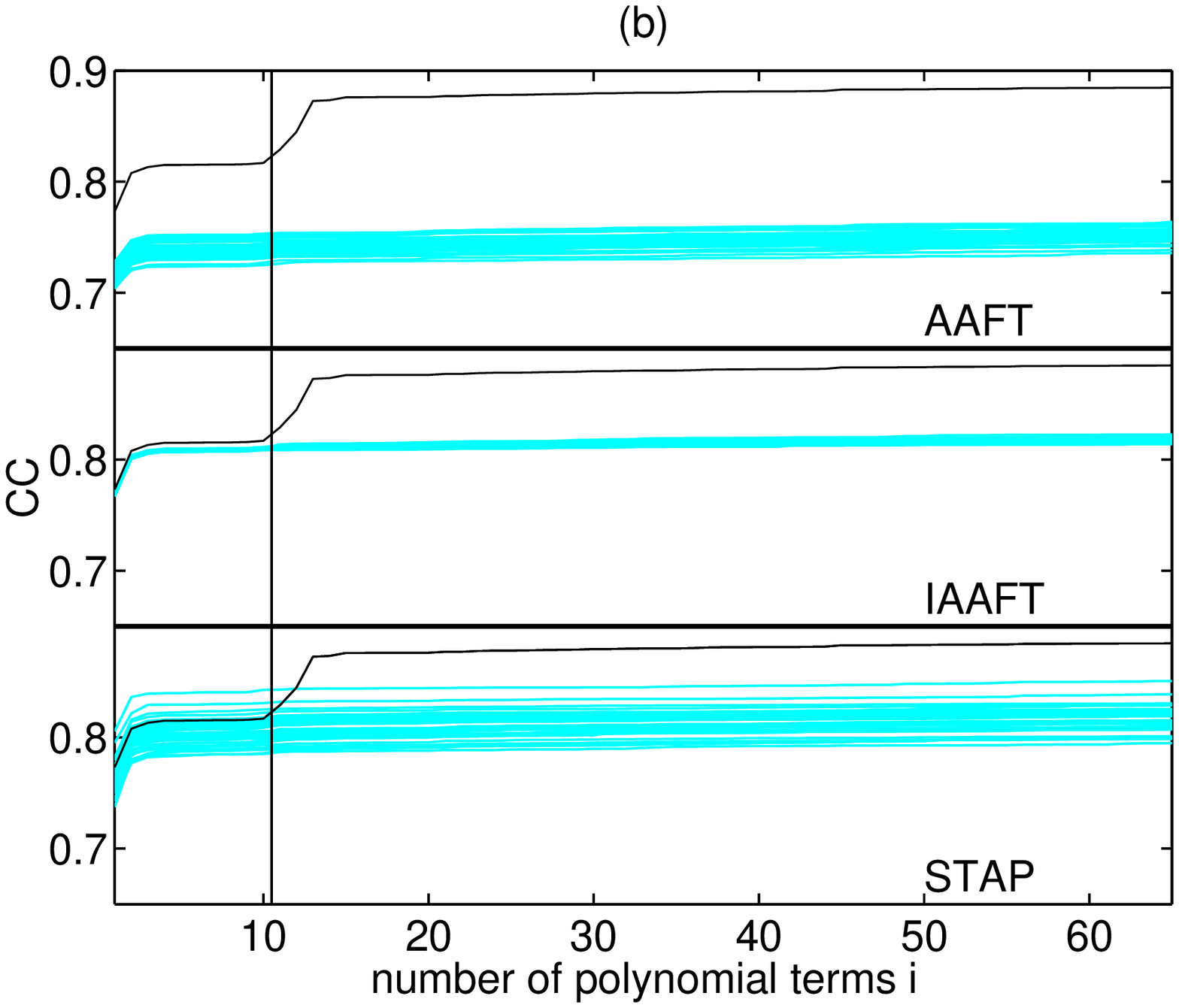}}
\centerline{\includegraphics[width=7cm,keepaspectratio]{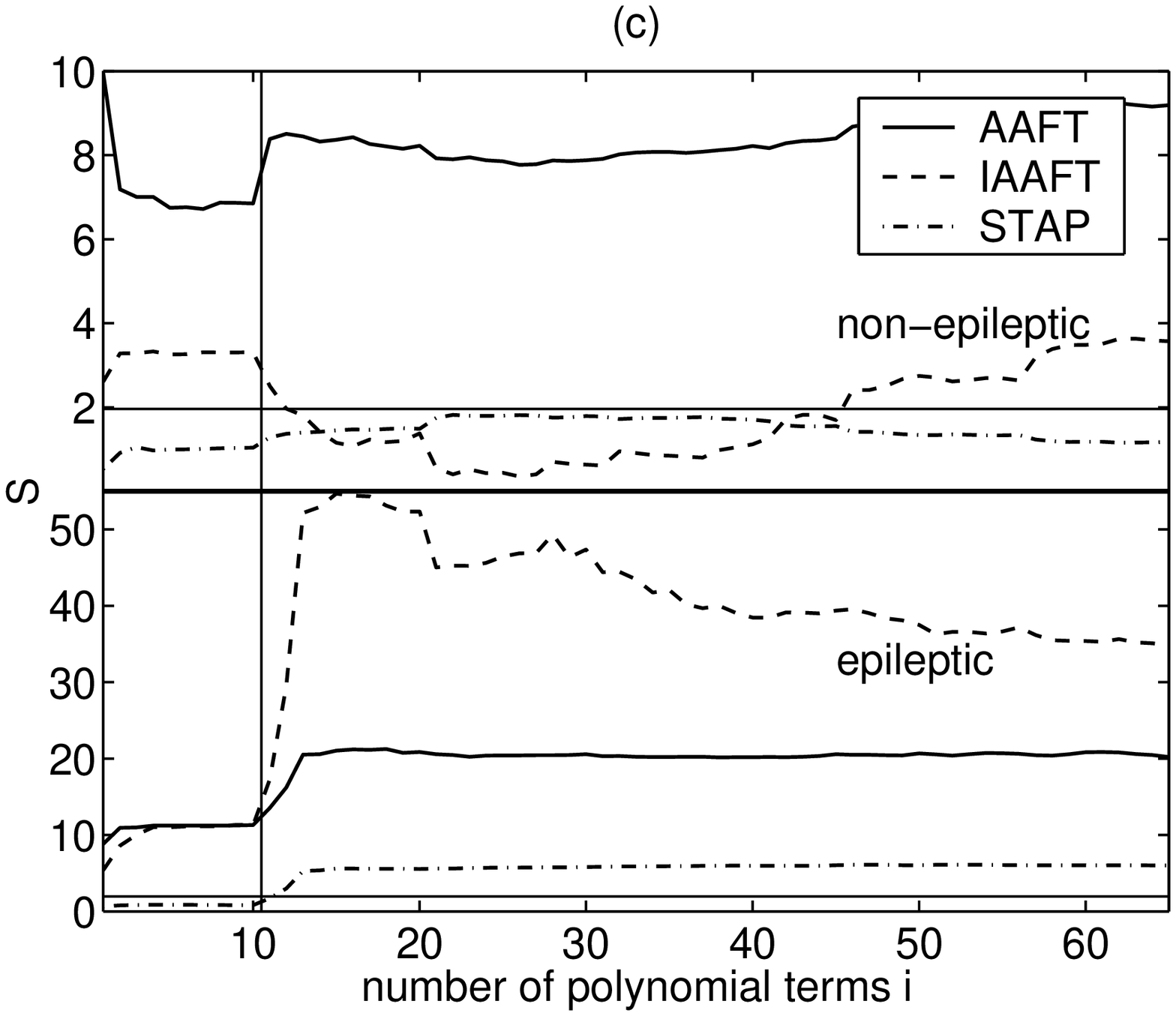}}
\caption{(a) The correlation coefficient of the fit with Volterra
  polynomials on the original normal EEG data (black line) and 
  40 AAFT, IAAFT and STAP surrogates (gray lines) in the three panels,
  as indicated (for STAP $m=10$, $p=10$). 
  (b) The same as in (a) but for the epileptic EEG.
  (c) The significance for the fits in (a) (upper panel) and in (b) 
  (lower panel) . 
  The vertical lines in the plots distinguish the linear from
  the nonlinear polynomials. 
}
\label{EEG}
\end{figure}

For the linear statistics, STAP gives consistently and correctly no 
rejections, i.e. unbiased match of the linear correlations, whereas
AAFT and IAAFT give very large percentages of rejections for all but
the first case where AAFT gives about $5\%$ rejections, as expected. 
For IAAFT, the rejections occur because $\sigma_q^i$ is very small 
($10$ to $20$ times smaller than for AAFT and STAP), though the bias 
$q_0^i - \langle q^i \rangle$ is smaller than for STAP.

For the nonlinear statistics, the feature of the linear statistics 
persists for the two first systems (consistent with $\mbox{H}_0$) and 
for all three algorithms, i.e. correctly no rejections with STAP and 
erronous rejections with AAFT and IAAFT.
We cannot explain why the polynomial fit for the IAAFT surrogates 
improves with the addition of nonlinear terms (see Fig.~\ref{Synthetic}).    
For the nonlinear system, STAP properly converges with the addition of 
few nonlinear terms to the correct $100\%$ rejection level, which AAFT
and IAAFT possessed already with linear statistics. 

Next, we verify the three algorithms on two human EEG data sets, one 
recorded many hours before an epileptic seizure accounting for normal 
brain activity and another recorded during an epileptic seizure.
The epileptic EEG seems to exhibit a pattern of oscillations
indicating a more deterministic character than the non-epileptic EEG,
and this constitutes a well established result in physiology. 
This is demonstrated also with the fit of Volterra polynomials
in Fig.~\ref{EEG}.
The fit improves with the inclusion of the first nonlinear terms for 
the epileptic EEG but not for the non-epileptic EEG. 
These findings are confirmed statistically by the test with STAP 
surrogate data while the results of the test with AAFT and IAAFT are 
more or less confusing. 

In particular, the $\mbox{H}_0$ on the normal EEG is erroneously
rejected with AAFT because the difference in the fit between original
and surrogate data is about the same for the linear and nonlinear
polynomials (see Fig.~\ref{EEG}a and c). 
The same test result is obtained with IAAFT for large nonlinear
polynomials, whereas again there is significant difference in the
linear fits between original and IAAFT surrogates (not easily
discernible as both bias and variance are very small).
Using STAP surrogates, the $\mbox{H}_0$ is not rejected for both
linear and nonlinear fits. 

For the epileptic EEG, there is again a clear difference in the 
linear fit between original data and AAFT surrogates and a smaller but 
equally significant difference between original data and IAAFT
surrogates (see Fig.~\ref{EEG}b and c). 
The significance $S$ for both AAFT and IAAFT increases with the 
addition of the first couple of nonlinear terms, much more for 
IAAFT due to the small variance of CC.   
However, the deviation in the linear fit does not support reliable
rejection of $\mbox{H}_0$.
On the other hand, using STAP surrogates, the $\mbox{H}_0$ is
properly rejected only for the nonlinear statistics and with high
confidence ($S<2$ for the linear fit and $S \simeq 5$ for the 
nonlinear fit). 

In general, the test with STAP surrogates tends to be more conservative,  
i.e. small discriminations are found less significant, as the data 
size decreases. 
For example, the test on 296 sunspot samples (for which a small leap 
of the polynomial fit with the addition of nonlinear terms was observed)
gave rejection of $\mbox{H}_0$ for AAFT and IAAFT but not for STAP 
(not shown here, see also \cite{Kugiumtzis00d}). 
However, this should not be considered as a drawback of the STAP
algorithm, as one expects that the power of the test reduces with the
decrease of data size. 

A new algorithm that generates surrogate data for the test for
nonlinearity has been presented, called statically transformed
autoregressive process (STAP). 
The key feature of STAP is that it finds analytically the
autocorrelation of an appropriate underlying normal process for the
test.  
This is the main difference of the STAP algorithm from the corrected
AAFT (CAAFT) algorithm, where the autocorrelation is estimated 
numerically.   
Both CAAFT and STAP algorithms do not suffer from the severe drawback
of the AAFT algorithm, i.e. bias in the match of the original
autocorrelation.
The AAFT algorithm is essentially impractical for real applications
because it favours the rejection of $\mbox{H}_0$ as a result of the
bias in the autocorrelation. 
From the numerical simulations, it turns out that the IAAFT algortihm
may also give small bias in the linear correlations, favouring also the
rejection of $\mbox{H}_0$.
On the other hand, the STAP algorithm performs properly and gives 
reliable rejections of $\mbox{H}_0$, only whenever this appears to be 
the case.

\end{document}